\documentclass[12pt,letterpaper]{article}
\usepackage[sort&compress, numbers]{natbib}
\usepackage{cite}
\usepackage{graphicx}
\usepackage{amssymb}
\usepackage[margin=1in,top=1in,bottom=1in]{geometry}
\usepackage{lineno}
\usepackage{enumitem}
\usepackage{aas_macros}
\usepackage{hyperref}
\usepackage{sectsty}
\sectionfont{\fontsize{15}{18}\selectfont}

\newcommand{\msun}{M_{\odot}}

\begin{document}

\begin{titlepage}
	\begin{centering}
	{\large\bfseries Astro2020 Science White Paper\par}
	\vspace{0.4cm}
	{\scshape\Large Fast Radio Burst Tomography \\ of the Unseen Universe \par}
	\vspace{0.4cm}
	
	\noindent {\mdseries \bf Thematic Areas:} {\em Cosmology and Fundamental Physics}, {\em Galaxy Evolution}  \\
	
	\vspace{0.4cm}

    { {\scshape Vikram Ravi}\footnote{Cahill Center for Astronomy and Astrophysics, MC\,249-17, California Institute of Technology, Pasadena CA 91125, USA; \href{mailto:vikram@caltech.edu}{vikram@caltech.edu}.}} {\normalsize\textit{(Center for Astrophysics $|$ Harvard \& Smithsonian; Caltech)}}; \\ 
    {  {\scshape Nicholas Battaglia}} {\normalsize\textit{(Cornell)}};
    {{\scshape Sarah Burke-Spolaor}} {\normalsize\textit{(West Virginia University/Center for Gravitational Waves and
Cosmology/CIFAR Azrieli Global Scholar)}};
    { {\scshape Shami Chatterjee}} {\normalsize\textit{(Cornell)}};
    {{\scshape James Cordes}} {\normalsize\textit{(Cornell)}};
    { {\scshape Gregg Hallinan}} {\normalsize\textit{(Caltech)}};
    { {\scshape Casey Law}} {\normalsize\textit{(UC Berkeley)}};
    { {\scshape T. Joseph W. Lazio}} {\normalsize\textit{(JPL/Caltech)}};
    { {\scshape Kiyoshi Masui}} {\normalsize\textit{(MIT)}};
    { {\scshape Matthew McQuinn}} {\normalsize\textit{(University of Washington)}};
    { {\scshape Julian B. Mu\~noz}} {\normalsize\textit{(Harvard University)}};
    { {\scshape Nipuni Palliyaguru}} {\normalsize\textit{(Arecibo Observatory)}};
    { {\scshape J. Xavier Prochaska}} {\normalsize\textit{(UC Santa Cruz)}};
    { {\scshape Andrew Seymour}} {\normalsize\textit{(Green Bank Observatory)}};
    { {\scshape Harish Vedantham}} {\normalsize\textit{(ASTRON)}};
    { {\scshape Yong Zheng}} {\normalsize\textit{(UC Berkeley)}}.

\vspace{0.4cm}

\end{centering}
\noindent {\mdseries \bf Description:} The discovery of Fast Radio Bursts (FRBs) at cosmological distances has opened a powerful window on otherwise unseen matter in the Universe. Observations of $>10^{4}$ FRBs will assess the baryon contents and physical conditions in the hot/diffuse circumgalactic, intracluster, and intergalactic medium, and test extant compact-object dark matter models. \\

\vspace{0.5cm}
	\begin{centering}
	\includegraphics[width=0.99\textwidth]{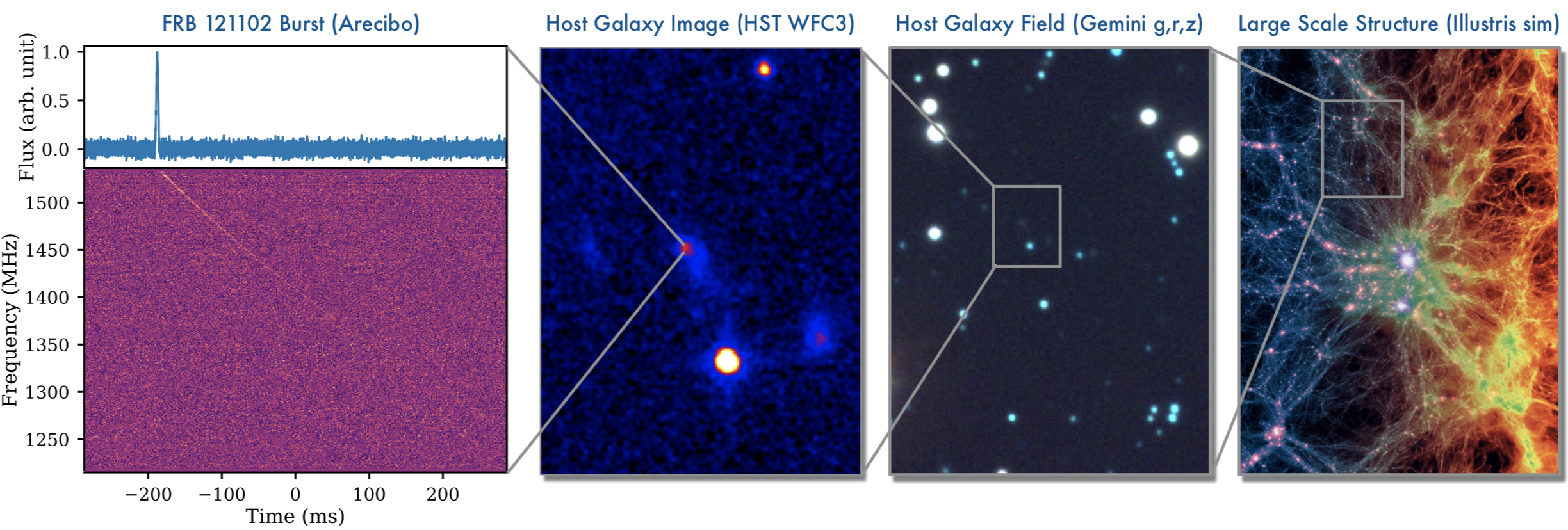}\par\vspace{0.2cm}
    \end{centering}
	{\small \emph{From left, Panel [1]}: A single burst from the repeating Fast Radio Burst (FRB) source 121102 detected at Arecibo, showing the frequency-dependent dispersion delay caused by the intervening electron column. [\emph{2}]: HST (F110W) image of the dwarf host-galaxy at a redshift $z=0.19273$, showing a star-forming knot associated with this burst source. [\emph{3}]: Gemini-N/GMOS image of the FRB\,121102 field, showing foreground stars and potentially intervening galaxies. [\emph{4}]: A simulation of the $z=0$ large-scale structure of the Universe on a 15\,Mpc/h scale, showing dark matter density on left, fading into the gas density on the right (Credit: The Illustris Simulation).} 

		\vfill
\end{titlepage}

Fast Radio Bursts (FRBs) are millisecond-duration impulses of extragalactic origin that offer a radically new means of observing the Universe \citep{lbm+07,tsb+13,pbj+16,r19}. At least $10^3$ events occur over the sky each day \citep{bkb+18,jem+19}, and a population of repeating sources is emerging \citep{ssh+16,cab+19}. FRBs are already detected on a daily basis in commensal searches with sensitive, widefield radio telescopes \citep{mbb+10,lbb+18,cab+18}. The next decade promises $10^4$--$10^6$ FRBs \citep{bkb+18,jem+19}, observed along unique extragalactic sightlines \citep{csr+19,j19} to redshifts $z\gtrsim2$ \citep{fl17,sd18,lrg+18}.  Here we describe how this bounty of FRBs will address three key questions:

\vspace{1mm}
\noindent \textbf{1. How are cosmic baryons in the low-redshift Universe allocated between galaxies, their surroundings, and the intergalactic medium (\S\ref{sec:2})?} 

\vspace{1mm}
\noindent \textbf{2. What do galaxy halos look like on the smallest scales? Are there parsec-scale cold clouds in the circum-galactic medium, and can even a fraction of dark matter be composed of compact objects (\S\ref{sec:3})?} 

\vspace{1mm}
\noindent \textbf{3. What are the physical conditions in the interstellar medium of galaxies besides the Milky Way (\S\ref{sec:4})? } 
\vspace{1mm}

    \begin{figure*}[b!]
    \center
    \includegraphics[width=0.75\textwidth]{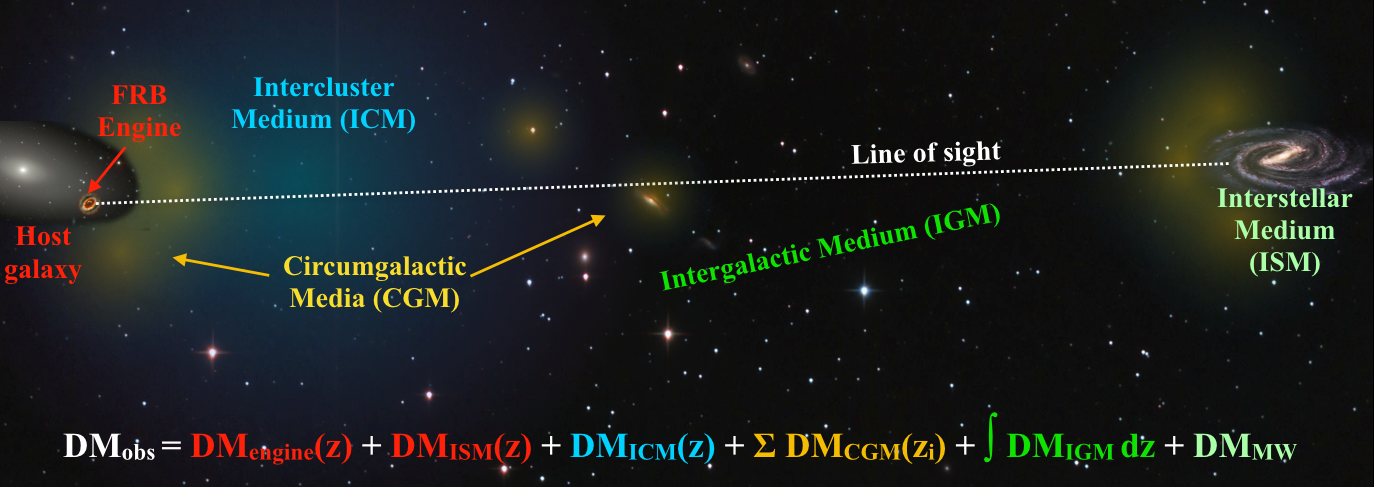}
    \vspace{-2ex}
        \caption{\small Illustration of the various proposed contributors to FRB DMs.}
        \label{fig:acronyms}
    \vspace{-3mm}
    \end{figure*}

The dispersive delays observed in all FRBs, estimated to sub-percent accuracy \citep[e.g.,][]{r19}, are significantly greater than expected from the Milky Way ionized interstellar and circumgalactic medium (ISM and CGM, respectively) \citep[see, e.g.,][]{kon+14}. FRB dispersion measures (DMs) quantify the total line-of-sight electron-column densities, and can include contributions from the host-galaxy ISM, the CGM and intra-cluster medium (ICM) of intervening systems, and the intergalactic medium (IGM) \citep[see Fig.~\ref{fig:acronyms};][]{i03,i04,m14,dgb+15,xh15,yz16,sd18,wmb18,pz19}. Just as radio pulsars help us model the Galactic ISM \citep{cwf+91,tc93,cl02,ymw17}, FRBs encode rich information on extragalactic warm/hot diffuse matter. Besides their DMs, FRBs provide the only means of measuring sub-ms time delays imparted by extragalactic lensing phenomena, which can be caused by both plasma-refractive and gravitational light bending. 

The scientific outcomes outlined in this white paper are largely independent of the nature of  FRB progenitors. Widespread interest in FRBs has so far been driven by the problem of their unknown origins -- e.g., the life cycles of their progenitors, and the mechanism of their emission \citep[for a catalog of FRB progenitor models, see][]{pww+18}. In the coming few years, several tens of FRBs will be localized with arcsecond accuracy \citep[e.g.,][]{mbb+10,lbb+18}. Although these data will significantly advance our understanding of FRB progenitors \citep[e.g.,][]{clw+17,tbc+17,mph+17}, ongoing FRB searches will likely be required to characterize FRB engines and explore their diversity. FRB survey data sets can also be used to search for other undiscovered, unexpected radio-astronomical phenomena.

\vspace{-5mm}
\section{The distribution of baryons in the low-$z$ universe} \label{sec:2}
\vspace{-2mm}

The Universe has been mostly ionized for the last 13 billion years \citep{m16}. Much of the IGM \citep[$\sim 50\%$ of the cosmic baryon density, $\Omega_{b}$;][]{ssd12} is now at $\gtrsim10^{7}$\,K, heated by shocks associated with structure formation and gas-accretion onto dark-matter halos \citep{co99}, and through thermal/kinetic feedback \citep[e.g.,][]{co06} from galaxies. Galactic outflows and IGM accretion also led to the establishment of the multi-phase CGM \citep{tpw17} and ICM \citep{rbn02}. The {\em missing baryon problem} refers to the difficulty of directly observing most cosmic baryons at redshifts $z\lesssim2$ \citep{fhp98,b07,ab10}. Approximately $0.3\Omega_{b}$ is in fact entirely absent from accounts of low-redshift baryons \citep{ssd12}. Recent observations of galactic coronae \citep{bam+18} and the thermal Sunyaev-Zeldovich (tSZ) effect in IGM filaments \citep{dch+17,thm+19}, and measurements of UV and X-ray quasar absorption lines \citep[e.g.,][]{gmk+12,wpt+14}, support a scenario wherein the missing baryons are hot ($>10^{6}$\,K) and diffusely distributed in the CGM, ICM, and IGM. 

\begin{figure*} 
    \center
    \includegraphics[width=0.99\textwidth]{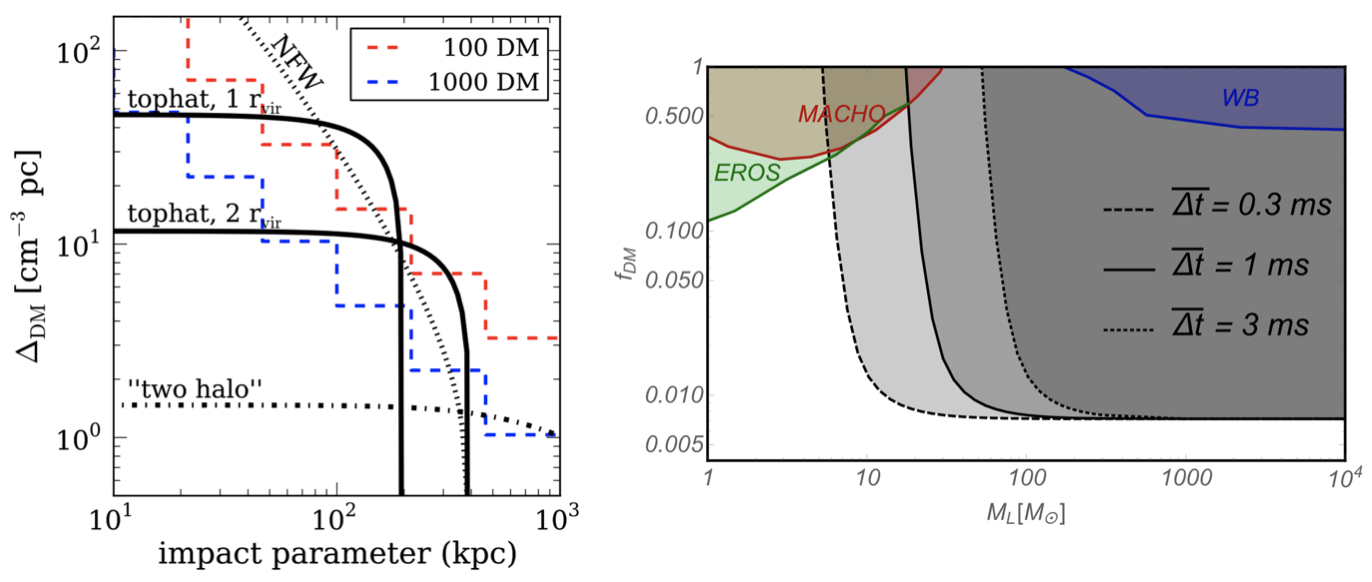}
\vspace{-3mm}
        \caption{\small {\em Left:} The sensitivity of different FRB sample sizes to the CGM in a $10^{12}\msun$ $z=0.5$ dark-matter halo \citep{m14}. The black curves show different models for the CGM density profile. The red and blue stepped lines show the $1\sigma$ uncertainties in the CGM DM, assuming $10^{2}$ and $10^{3}$ FRBs {\em per bin} respectively.  {\em Right:} The constraints on the fraction of dark matter ($f_{\rm DM}$) in primordial black holes for different masses ($M_{L}$) if none out of $10^{4}$ FRBs are micro- or nano-lensed (grey) \citep{mkd+16}. A set of current constraints is shown as red, green and blue regions.}
        \label{fig:1}
\vspace{-3mm}
\end{figure*}

Large FRB/DM samples will enable a direct accounting of the $z\lesssim2$ baryon contents of the CGM, ICM, and IGM respectively, without detailed cosmological and photo-chemical simulations \citep[e.g.,][]{wpt+14,msb+17,osc+18} to interpret the observations. Just $10^{1}-10^{2}$ arcsecond-localized FRBs, with host- and intervening-system redshift measurements, are sufficient to make a statistical detection of the CGM and measure its corresponding fraction of $\Omega_{b}$ \citep{r19b}.  Stacking $10^{2}-10^{3}$ arcminute-localized FRBs in individual bins of impact parameter to intervening galaxies will additionally probe the radial density profile of the CGM \citep[Fig.~\ref{fig:1};][]{m14}. Larger FRB samples will measure the CGM electron-column density profiles for different galaxy types, addressing a major uncertainty in interpreting observations based on UV metal-absorption lines and X-ray emission measures \citep[e.g.,][]{ab10,ssd12,wpt+14}. Further, the correlation of $\sim10^{4}$ arcminute-localized FRBs and their DMs with cosmic microwave background (CMB) maps will measure the electron-scattering optical depths corresponding to the cosmic-web tSZ effect at higher redshifts than are currently possible \citep{mu18}. Finally, the measurement of the DM-$z$ relation to $z>2$ with samples of $\sim10^{4}$ FRBs to mitigate cosmic variance will ascertain the precise epoch and duration of He\,\textsc{ii} reionization at $z\sim2$ \citep{zok+14,cfs19}.

Although $\gg10^{4}$ FRB DMs with associated redshifts are required to improve estimates of cosmological parameters \citep{zlw+14,wwg+18,j19}, three important cosmological measurements are more readily achievable. The DM-space clustering of $\sim10^{4}-10^{5}$ arcminute-localized events will probe large-scale structure through baryon acoustic oscillations, and characterize the bias parameter of cosmic baryons \citep{ms15}.  Further, FRBs can be used to improve galaxy-cluster kinetic SZ (kSZ) measurements of the growth rate and amplitude of cosmic density fluctuations \citep{MatFRB2019}. This is an important goal of Stage-4 CMB experiments, and the Dark Energy Spectroscopic Instrument (DESI) surveys. KSZ constraints are degenerate with galaxy-cluster CMB optical depths \citep[e.g.,][]{Bat2016,FNM2017}, which are currently calibrated using cosmological simulations. $10^{5}-10^{6}$ arcsecond-localized FRBs with even photometric-redshift information, jointly with DESI survey data, can break this degeneracy and improve kSZ constraints by factors of $\sim2$ \citep{MatFRB2019}. Finally, the DMs and Faraday rotation measures (RMs -- which are the sightline-averaged products of the electron density and magnetic field) of linearly polarized FRBs enable uniquely direct estimates of mean line-of-sight extragalactic magnetic fields \citep[e.g.,][]{mls+15,rsb+16}. A handful of FRB RMs have already been measured \citep{pbj+16}, and $\gtrsim10^{3}$ FRBs with extragalactic RM uncertainties of a few rad\,m$^{-2}$ will distinguish between leading models for seeding the as-yet undetected IGM magnetic field \citep{arg16,vbh+18}.

\vspace{-5mm}
\section{Extragalactic sub-microlensing of FRBs} \label{sec:3}
\vspace{-2mm}

FRBs are the shortest-duration extragalactic transients, and the most compact known extragalactic sources of electromagnetic radiation. FRBs hence resolve smaller time-delays caused by multi-path propagation on cosmological scales than any other probe, due to gravitational and plasma-refractive light bending. FRBs refracted in inhomogeneous plasma arrive at the observer along several paths, and propagation along gravitationally deflected paths likewise enables multiple burst copies to be detected. Few-microsecond temporal structure exists in some FRBs \citep{rsb+16,ffb+18}, implying sensitivity to comparable propagation delays in total-power time series. Time delays, $\tau$, comparable to the inverse of the observing bandwidths ($\sim 1$~ns) can be detected through the (frequency-dependent) interference of differently delayed rays. The corresponding angular scale at the plasma or gravitational ``lens'', at a distance $D_{l}$, is $\theta_{l}=\sqrt{c\tau/D_{l}}\approx\sqrt{[\tau/(1\,\mu{\rm s})][(1\,{\rm Gpc}/D_{l}]}$\,${\rm \mu}$as. Such angular scales are difficult to spatially resolve with even mm Very Long Baseline Interferometry (VLBI), and transient lensing events are inaccessibly long except in special cases \citep{kd00}, {\em making FRB observations the best means to detect extragalactic micro- and nano-lenses}. Gravitational lensing is achromatic, unlike plasma-refractive lensing, and lensing unlike intrinsic temporal structure imposes a strong temporal coherence on the radiation.

A major, unexpected challenge to our understanding of the CGM \citep[e.g.,][]{bd03,kkw+05,csv+18} is the ubiquitous detection of cool ($\sim10^{4}$\,K), dense ($\sim1$\,cm$^{-3}$) material admixed with $10^{5}-10^{7}$\,K gas \citep{tpw17}. The cool CGM gas could originate from galactic outflows \citep{hfa+18,sng+19}, or cooling instabilities may be amplified by magnetic fields \citep{jom18}. Alternatively, the gas may cool in situ on scales below simulation resolutions, fragmenting into a fog of cold clumps ($\sim0.1$\,pc, $\sim1$\,cm$^{-3}$) \citep{moo+18}. These clumps are likely only detectable through the resulting `scattering' (stochastic multi-path propagation) of FRB pulses by CGM halos on timescales of $\sim(\nu/{\rm GHz})^{-4}$\,ms \citep{vp19}. More halo intercepts will result in increased scattering; FRBs at $z>1$ will intersect $\sim10$ Milky Way-sized halos. Such chromatic scattering has long been observed in Galactic pulsars due to the Milky Way ISM \citep[e.g.,][]{cwf+91}, and is seen in several FRBs \citep{pbj+16,cws+16,r19}. The FRB sample in hand and experience with Galactic pulsars together suggests that establishing a scattering-redshift relation will require well over $10^{3}$ (arcsecond-localized)  FRBs. 

Gravitational lensing is a classic probe of unseen matter in the Universe, and FRB time-delay observations can probe extragalactic mass concentrations on otherwise inaccessible scales. 
If the entirety of cosmological dark matter is composed of compact objects, such as $10-100\msun$ black holes (that are not probed by the usual stellar-microlensing searches) \citep[e.g.,][]{bcm:2016,cks16,crt+17}, lensed FRB echoes on $10^{-6}-10^{-3}$\,s timescales will exist for $\sim1/100$ events \citep[Fig~\ref{fig:1};][]{mkd+16}. Gravitational-lensing searches with $10^{4}$ FRBs will assess whether even $1\%$ of dark matter consists of compact objects \citep{mkd+16,e17,l18}. Finally, repeating FRBs can be used to probe temporal changes in larger-scale lensing potential time delays with unprecedented accuracy, providing the most promising means to directly observe transverse motion on cosmological scales, and even the Hubble expansion of the Universe \citep{lgd+18,ze18,wle19}.

\vspace{-5mm}
\section{FRB sightlines through extragalactic (and Galactic) ISM} \label{sec:4}
\vspace{-2mm}

The remarkable scientific outcomes forecast above first require the characterization of FRB-propagation effects in their host galaxies, and in the Milky Way. This presents a unique opportunity to study extragalactic ISM. 

Known FRBs are viewed along a wide variety of plasma sightlines. FRB\,150807, for example, likely has a small host-galaxy DM, together with an undetectable extragalactic RM, and a scattering delay of $5-10$\,$\mu$s \citep{rsb+16}. The repeating FRB\,121102 has a host-galaxy DM of $\sim200$\,pc\,cm$^{-3}$ (comparable to its IGM DM at $z=0.19273$) \citep{tbc+17}, among the highest RMs ever observed among radio sources \citep{msh+18}, but with only Galactic-ISM scattering \citep{mph+17}. FRBs with measurable extragalactic scattering appear not to be scattered in environments typical of the Milky Way ISM \citep{mls+15,rsb+16,cab+19b}. Although the predictions made above for the utility of FRB DMs account for realistic uncertainties estimated for host- and Galactic-DM contributions, the former can be assessed through joint analyses of FRB DMs, scattering measurements, and RMs \citep{rsb+16,cws+16,cab+19b}. For FRBs where Milky Way scattering is identified, the locations of additional scattering instances (i.e., the CGM, or the host galaxies) can be deduced through geometric-optics arguments \citep{mls+15,ffb+18}. With large FRB samples, it will become possible to identify the DMs, RMs, and plasma-refractive scattering phenomena contributed by different FRB host-galaxy types. Repeated bursts from some FRBs can be used to monitor the host-galaxy ISM properties on timescales of years, probing AU-scale density inhomogeneities in extragalactic ISM. DM and RM variations in the repeating FRB\,121102 have indeed already been reported \citep{hss+18}, with important consequences for the progenitor environment \citep{vr18}.

Models of the density distribution and physical conditions of foreground Galactic ionized ISM \citep[e.g.,][]{cl02,ymw17} will be improved by factors of a few by next-generation pulsar surveys \citep[yielding $\sim10^{4}$ discoveries;][]{b18,lab+18}, and more VLBI parallax measurements of pulsars. The next generation of polarimetric radio-continuum sky surveys will lead to similarly improved models for the Galactic RM foreground \citep[e.g.,][]{ojg+15}. $\gtrsim10^{4}$ FRB sightlines through the Galactic halo can in turn address the factor of $\sim2$ uncertainty in its total baryon content \citep{wmb18,pz19}.

\vspace{-5mm}
\section{Prospects and requirements for the 2020s} \label{sec:5}
\vspace{-2mm}

At least $10^{3}$ FRBs occur over the sky each day, and the boundaries of the parameter space of FRB properties remain unknown. Current telescopes can achieve detection rates of a few per day. A sample of $10^{4}-10^{6}$ events, with a substantial repeating fraction, is therefore expected over the coming decade. Evidence from FRB sky-localization regions \citep{rsb+16,clw+17}, analyses of the FRB fluence distribution \citep{vrh+16,me18,jem+19}, and a nascent FRB fluence -- DM correlation \citep{smb+18} together evince a population of events  at cosmologically significant distances. Although the FRB engines are currently unknown, the merits of $\gtrsim10^{4}$  extragalactic sightlines along which matter column-densities can be measured are manifest. FRBs are additionally the best probes of extragalactic plasma inhomogeneities and massive compact objects that result in sub-millisecond multi-path delays. {\bf Scientific possibilities of various FRB samples:}

\vspace{1mm}
\noindent \textbf{$\mathbf{10^{3}-10^{4}}$ FRBs.} Statistical detections of CGM gas densities at different impact parameters to intervening galaxies ($\lesssim10^{\prime\prime}$\ localizations) --- characterization (if present) of a scattering-redshift relation, testing CGM-cooling models ($\lesssim1^{\prime\prime}$ localizations) --- ascertaining if all of dark matter is composed of compact objects. 

\vspace{1mm}
\noindent \textbf{$\mathbf{10^{4}-10^{5}}$ FRBs.} Measurements of the mean IGM density and magnetic field in the cosmic web ($\lesssim60^{\prime\prime}$ localizations) --- global detection of He\,\textsc{ii} reionization ($\lesssim1^{\prime\prime}$ localizations) --- detection of baryon acoustic oscillations in DM-space clustering ($\lesssim60^{\prime\prime}$ localizations).

\vspace{1mm}
\noindent \textbf{$\mathbf{>10^{5}}$ FRBs.} Improve kSZ constraints on large-scale structure growth rate ($\lesssim3^{\prime\prime}$ localizations) --- detect extragalactic baryonic gravitational micro- and nano-lenses. 
\vspace{1mm}

\noindent These outcomes may rely on the assembly of carefully controlled subsamples of FRBs, much as subsamples of galaxies are used for cosmological applications. This may require total FRB samples that are up to an order of magnitude larger. {\em The ideal FRB-detection radio telescope for the next decade must occupy the intersection of moderate sensitivity, arcsecond localization capability, and an immense (several tens of \,deg$^{2}$) field of view.}

These scientific milestones require the synthesis of additional inputs. Improved models for the Milky Way ionized-ISM distribution and physical conditions will be important for foreground characterization in DM, RM, and scattering measurements of FRBs. The characterization of FRB host galaxies may require substantial pointed optical/infrared spectroscopic observations \citep[e.g.,][]{tbc+17,eb17}, although large-scale photometric redshift catalogs (e.g., from the Large Synoptic Survey Telescope) will prove sufficient in several cases \citep[e.g.,][]{MatFRB2019}. Multiplexed spectroscopic surveys (e.g., with DESI) covering much of the sky are  required in addition to optical imaging surveys to identify intervening systems along FRB sightlines. 

The studies described here elegantly complement and synergize with multi-wavelength probes of unseen matter. Optical / UV / X-ray absorption-line studies of the CGM and IGM often probe small gas fractions at $z\lesssim2$ (by temperature and composition), whereas FRB DMs correspond to the bulk gas contents. Thermal X-ray observations of the CGM and ICM are largely sensitive to the densest parts of the gas distributions, because of the density-squared dependence of the emission measure. Stellar microlensing searches for unseen massive compact objects are most efficacious in the Milky Way, whereas FRB lensing studies (like ground-based gravitational-wave searches) are more sensitive to these systems at extragalactic distances. FRBs are hence poised to become a leading means of characterizing the unseen matter of the Universe. 

\clearpage

\bibliographystyle{unsrt}
\bibliography{bib.bib}

\end{document}